\documentclass[12pt]{article}
\usepackage{graphicx}
\usepackage{amsmath}
\usepackage{amssymb}
\usepackage{cite}
\usepackage{multirow}
\usepackage{framed}
\usepackage{slashed}
\usepackage{bbm}
\usepackage{array}

\newcount\hour\newcount\minute
        \hour=\time \divide\hour by60 \minute=\time
        {\multiply\hour by60 \global\advance\minute by-\hour}
        \edef\militarytime{\number\hour:\ifnum\minute<10
0\fi\number\minute}

\makeatletter
\@addtoreset{equation}{section}

\def\asymp#1%
{\mathrel{\raisebox{-.4em}{$\widetilde{\scriptstyle #1}$}}}
\makeatother

\newcommand\Ref[1] {Ref.\,\cite{#1}}
\newcommand\Refs[1] {Refs.\,\cite{#1}}
\newcommand\eqn[1] {Eq.\,(\ref{#1})}

\newcommand\fig[1] {Fig.\,{\ref{#1}}}
\newcommand\figs[2] {Figs.\,(\ref{#1}) and~(\ref{#2})}

\newcommand\app[1] {Appendix~\ref{#1}}
\newcommand\tab[1] {Table~\ref{#1}}

\newcommand\subtitle[1] {\noindent{\bf #1}}

\def\beq{\begin{equation}}
\def\eeq{\end{equation}}
\def\bsp#1\esp{\begin{split}#1\end{split}}
\def\bal#1\eal{\begin{align}#1\end{align}}
\def\beeq{\begin{eqnarray}}
\def\eeeq{\end{eqnarray}}

\newcommand\bom[1]     {{\mbox{\boldmath $#1$}}}
\newcommand\lp   {\ensuremath{\left}}
\newcommand\rp   {\ensuremath{\right}}

\newcommand\rb   {\ensuremath{\mathrm{b}}}
\newcommand\rc   {\ensuremath{\mathrm{c}}}
\newcommand\rd   {\ensuremath{\mathrm{d}}}
\newcommand\re   {\ensuremath{\mathrm{e}}}
\newcommand\ri   {\ensuremath{\mathrm{i}}}
\newcommand\rt   {\ensuremath{\mathrm{t}}}
\newcommand\rL   {{\mathrm{L}}}
\newcommand\rR   {{\mathrm{R}}}

\newcommand\rY   {{\mathrm{Y}}}

\newcommand\GeV  {{\mathrm{GeV}}}

\newcommand\tS   {\theta_S}
\newcommand\tZ   {\theta_{Z}}

\newcommand\bT   {\bom{T}}

\newcommand\cL   {\ensuremath{\mathcal{L}}}

\newcommand\sm   {\ensuremath{\mathrm{SM}}}
\newcommand\lSM  {\ensuremath{\lambda_\sm}}
\newcommand\mt   {\ensuremath{m_{\rt}}}
\newcommand\mP   {\ensuremath{m_{\mathrm{P}}}}

\def\draftdate{\relax}
\def\mda{\relax}
\def\mua{\relax}
\def\mla{\relax}
\def\draft{
\def\thtystars{******************************}
\def\sixtystars{\thtystars\thtystars}
\typeout{}
\typeout{\sixtystars**}
\typeout{* Draft mode!
For final version remove \protect\draft\space in source file *}
\typeout{\sixtystars**}
\typeout{}
\def\draftdate{\today}
\def\mua{\marginpar[\boldmath\hfil$\uparrow$]%
{\boldmath$\uparrow$\hfil}%
\typeout{marginpar: $\uparrow$}\ignorespaces}
\def\mda{\marginpar[\boldmath\hfil$\downarrow$]%
{\boldmath$\downarrow$\hfil}%
\typeout{marginpar: $\downarrow$}\ignorespaces}
\def\mla{\marginpar[\boldmath\hfil$\rightarrow$]%
{\boldmath$\leftarrow $\hfil}%
\typeout{marginpar: $\leftrightarrow$}\ignorespaces}
\overfullrule 5pt
\oddsidemargin -15mm
\marginparwidth 29mm
}

\def\stars{\strut\leaders\hbox{*}\hfill\strut}
\def\starline{\hfil\strut\hfil\hbox to \textwidth {\stars}\hfil}

\begin{document}

%%%%%%%%%%%%%%%%%%%%%%%%%%%%%%%%%%%%%%%%%%%%%%%%%%%%%%%%%%%%%%%%%%%%%%
%%%%%%%%%%%%%%%%%%%%%%%%%%%%%%%%%%%%%%%%%%%%%%%%%%%%%%%%%%%%%%%%%%%%%%
%%% 
%%% Title
%%%
%%%%%%%%%%%%%%%%%%%%%%%%%%%%%%%%%%%%%%%%%%%%%%%%%%%%%%%%%%%%%%%%%%%%%%
%%%%%%%%%%%%%%%%%%%%%%%%%%%%%%%%%%%%%%%%%%%%%%%%%%%%%%%%%%%%%%%%%%%%%%

\begin{titlepage}
\renewcommand{\thefootnote}{\fnsymbol{footnote}}
\begin{flushright}
arXiv:yymm.nnnn 
\end{flushright}
\par \vspace{5mm}
\begin{center}
{\Large \bf
Stability of the vacuum as constraint on $U$(1) extensions of the 
standard model
}
\end{center}
\par \vspace{2mm}

\begin{center}
%\author{Zolt\'an Tr\'ocs\'anyi}
{\bf Zolt\'an P\'eli$^a$
and Zolt\'an Tr\'ocs\'anyi$^{a,b}$},\\[.5em]
%{Affiliation
%}
{\it $^a$MTA-DE Particle Physics Research Group,\\
H-4010 Debrecen, PO Box 105, Hungary\\
$^b$Institute for Theoretical Physics, E\"otv\"os Lor\'and University, \\
P\'azm\'any P\'eter s\'et\'any 1/A, H-1117 Budapest, Hungary 
}\\
E-mail: {Zoltan.Trocsanyi@cern.ch}
\end{center}

\par \vspace{2mm}
\begin{center}
\today
\end{center}

\par \vspace{2mm}
\begin{center} {\large \bf Abstract} \end{center}
\begin{quote}
\pretolerance 10000
In the standard model the running quartic coupling becomes negative
during its renormalization group flow, which destabilizes the vacuum.
We consider U(1) extensions of the standard model, with an extra
complex scalar field and a Majorana-type neutrino Yukawa coupling.
These additional couplings affect the renormalization group flow of the
quartic couplings. We compute the beta-functions of the extended model
at one-loop order in perturbation theory and study how the parameter
space of the new scalar couplings can be constrained by the requirement
of stable vacuum and perturbativity up to the Planck scale.
\end{quote}

\vspace*{\fill}
\begin{flushleft}
2019
\end{flushleft}
\end{titlepage}
\clearpage

\renewcommand{\thefootnote}{\fnsymbol{footnote}}

%%%%%%%%%%%%%%%%%%%%%%%%%%%%%%%%%%%%%%%%%%%%%%%%%%%%%%%%%%%%%%%%%%%%%%
%%%%%%%%%%%%%%%%%%%%%%%%%%%%%%%%%%%%%%%%%%%%%%%%%%%%%%%%%%%%%%%%%%%%%%
%%%
%%% Introduction
%%%
%%%%%%%%%%%%%%%%%%%%%%%%%%%%%%%%%%%%%%%%%%%%%%%%%%%%%%%%%%%%%%%%%%%%%%
%%%%%%%%%%%%%%%%%%%%%%%%%%%%%%%%%%%%%%%%%%%%%%%%%%%%%%%%%%%%%%%%%%%%%%

%\section{Introduction}
%\label{sec:intro}

The standard model of elementary particle interactions \cite{Weinberg:1967tq}
has been proven experimentally to high precision at the Large Electron
Positrion Collider \cite{ALEPH:2005ab} and also at the Large Hadron
Collider (LHC) \cite{ATLAS,CMS}.  At the LHC the last missing piece,
the Higgs particle has also been discovered and its mass has been
measured at high precision \cite{Aad:2014aba,Khachatryan:2014jba}, which
made possible the precise renormalization group (RG) flow analysis of
the Brout-Englert-Higgs potential \cite{Degrassi:2012ry,Buttazzo:2013uya}.
The perturbative precision of this computation is sufficiently high so
that the conclusion about the instability of the vacuum in the standard
model cannot be questioned. While the instability may not influence the
fate of our present Universe if the tunneling rate from the false vacuum is
sufficiently low (making the Universe metastable), one may insist that
the vacuum must be stable up to the Planck scale. Indeed, if we assume
the natural proposition that cosmological time is inversely
proportional to the relevant energy scale of particle processes, then
short after the Big Bang the Universe based on the standard model were
unstable and could not exist, which calls for an extension of the
standard model.

In this letter we consider the simplest possible extension of the 
standard model gauge group $G_\sm = SU(3)_\rc\otimes SU(2)_\rL\otimes
U(1)_Y$ to $G_\sm \otimes U(1)_Z$ and study the renormalization group
flow of the scalar couplings at one-loop order in perturbation theory.
Although, we are motivated by a specific model of such extensions
\cite{Trocsanyi:2018bkm}, for small values of the new gauge couplings--as
suggested by other phenomenological considerations--the only relevant
couplings are the scalar ones and the largest Yukawa-coupling in the
neutrino sector if we assume similar hierarchy of the latter as one can
observe for u-type quarks in the standard model \cite{Tanabashi:2018oca}.
Hence, the precise formulation of the gauge sector does not influence our
conclusions and we need to focus on the formulation of the scalar sector.

%%%%%%%%%%%%%%%%%%%%%%%%%%%%%%%%%%%%%%%%%%%%%%%%%%%%%%%%%%%%%%%%%%%%%%
%%%%%%%%%%%%%%%%%%%%%%%%%%%%%%%%%%%%%%%%%%%%%%%%%%%%%%%%%%%%%%%%%%%%%%
%%%
%%% Scalar sector
%%%
%%%%%%%%%%%%%%%%%%%%%%%%%%%%%%%%%%%%%%%%%%%%%%%%%%%%%%%%%%%%%%%%%%%%%%
%%%%%%%%%%%%%%%%%%%%%%%%%%%%%%%%%%%%%%%%%%%%%%%%%%%%%%%%%%%%%%%%%%%%%%

Our scalar sector is defined similarly as in the standard model, but in
addition to the usual scalar field $\phi$ that is an $SU(2)_\rL$-doublet
\begin{equation}
\phi=\lp(\!\!\begin{array}{c}
               \phi^{+} \\
               \phi^{0}
             \end{array}\!\!\rp) = 
\frac{1}{\sqrt{2}}
\lp(\!\!\begin{array}{c}
          \phi_{1}+\ri\phi_{2} \\
          \phi_{3}+\ri\phi_{4}
        \end{array}\!\!
\rp)
\,,
\end{equation}
there is also another complex scalar $\chi$ that transforms as a
singlet under $G_{\rm SM}$ transformations.  The gauge invariant
Lagrangian of the scalar fields is
\beq
\cL_{\phi,\chi} =
  [D^{(\phi)}_\mu \phi]^* D^{(\phi)\,\mu} \phi
+ [D^{(\chi)}_\mu \chi]^* D^{(\chi)\,\mu} \chi
- V(\phi,\chi)
\,.
\label{eq:Lphichi}
\eeq
The covariant derivative for the scalar $s$ ($s=\phi$, $\chi$) is
\beq
D_{\mu}^{\lp(s\rp)}= \partial_{\mu}
+\ri g_\rL\,\bT\cdot\bom{W}_{\mu}
+\ri\,y_{s} g_Y B'_{\mu}
+\ri \lp(r_s g'_Z + y_{s} g'_{ZY} \rp)Z'_{\mu}
\label{eq:cov-dev-scalar}
\eeq
where $\bT = (T^1,T^2,T^3)$ are the generators and $g_\rL$ is the
coupling of the $SU(2)_\rL$ group, $g_Y$ is the $U(1)_Y$ coupling,
$g'_Z=g_Z/\cos\tZ$ is the ratio of the $U(1)_Z$ coupling and the cosine
of the kinetic mixing angle and $g'_{ZY}=g'_Z - g_Y \tan\tZ$ is the
mixed coupling \cite{delAguila:1995rb}, while $y_s$, $r_s$ are the
corresponding hyper- and super-weak charges of the scalars. In the
renormalization group analysis below we shall concentrate on the
phenomenologically relevant case when the new couplings are super-weak,
hence negligible in the scalar sector, and so the actual values of
$r_s$ are irrelevant.  

In \eqn{eq:Lphichi} the potential energy
\beq
V(\phi,\chi) = V_0 - \mu_\phi^2 |\phi|^2 - \mu_\chi^2 |\chi|^2
+ \lp(|\phi|^2, |\chi|^2\rp)
\lp(\!\!\begin{array}{cc}
 \lambda_\phi & \frac{\lambda}2 \\ \frac{\lambda}2 & \lambda_\chi 
\end{array}\!\!\rp) 
\lp(\!\!\begin{array}{c}
|\phi|^2 \\ |\chi|^2
\end{array}\!\!\rp)\,,
\label{eq:V}
\eeq
in addition to the usual quartic terms, introduces a coupling term
$-\lambda |\phi|^2 |\chi|^2$ of the scalar fields in the Lagrangian
where $|\phi|^2 = |\phi^+|^2 + |\phi^0|^2$.
The value of the additive constant $V_0$ is irrelevant for particle
dynamics, but may be relevant for inflationary scenarios, hence we
allow a non-vanishing value for it. In order that this potential energy
be bounded from below, we have to require the positivity of the
self-couplings, $\lambda_\phi$, $\lambda_\chi>0$. The eigenvalues
of the coupling matrix are
\beq
\lambda_\pm = \frac12 \left(\lambda_\phi+\lambda_\chi
\pm \sqrt{(\lambda_\phi-\lambda_\chi)^2 + \lambda^2}\right)
\,,
\eeq
with $\lambda_+>0$ and $\lambda_-<\lambda_+$. In the physical region the
potential can be unbounded from below only if $\lambda_-<0$ and the
eigenvector belonging to $\lambda_-$ points into the first quadrant,
which may occur only when $\lambda<0$. In this case, the potential will
be bounded from below if the coupling matrix is positive definite, i.e.
\beq
4 \lambda_\phi \lambda_\chi - \lambda^2 > 0
\,.
\label{eq:positivity}
\eeq
If these conditions are satisfied, we find the minimum of the
potential energy at field values $\phi= v/\sqrt{2}$ and
$\chi = w/\sqrt{2}$ where the vacuum expectation values (VEVs) are
\beq
v = \sqrt{2} \sqrt{\frac{2\lambda_\chi \mu_\phi^2 - \lambda \mu_\chi^2}
{4 \lambda_\phi \lambda_\chi - \lambda^2}}
\,,\qquad
w = \sqrt{2} \sqrt{\frac{2\lambda_\phi \mu_\chi^2 - \lambda \mu_\phi^2}
{4 \lambda_\phi \lambda_\chi - \lambda^2}}
\,.
\label{eq:VEVs}
\eeq
Using the VEVs, we can express the quadratic couplings as
\beq
\mu_\phi^2 = \lambda_\phi v^2 + \frac{\lambda}{2} w^2
\,,\qquad
\mu_\chi^2 = \lambda_\chi w^2 + \frac{\lambda}{2} v^2
\,,
\label{eq:scalarmasses}
\eeq
so those are both positive if $\lambda > 0$. If $\lambda < 0$, the
constraint (\ref{eq:positivity}) ensures that the denominators of the
VEVs in \eqn{eq:VEVs} are positive, so the VEVs have non-vanishing real
values only if
\beq
2\lambda_\chi \mu_\phi^2 - \lambda \mu_\chi^2 > 0
\quad\text{and}\quad
2\lambda_\phi \mu_\chi^2 - \lambda \mu_\phi^2 > 0
\label{eq:muXconditions}
\eeq
simultaneously, which can be satisfied if at most one of the quadratic
couplings is smaller than zero. We summarize the possible cases for the
signs of the couplings in \tab{tab:scalarcouplings}.
\begin{table}
\begin{center}
\caption{Possible signs of the couplings in the scalar potential 
$V(\phi,\chi)$ in order to have two non-vanishing real VEVs.
$\Theta$ is the step function, $\Theta(x)=1$ if $x>0$ and 0 if $x<0$}
\label{tab:scalarcouplings}
\begin{tabular}{c|ccccc}
\hline
\hline
$\Theta(\lambda)$ &
$\Theta(\lambda_\phi)$ & $\Theta(\lambda_\chi)$ &
$\Theta(4 \lambda_\phi \lambda_\chi - \lambda^2)$ &
$\Theta(\mu_\phi^2)\,\Theta(\mu_\chi^2)$ &
$\Theta(2\lambda_\chi \mu_\phi^2 - \lambda \mu_\chi^2)
 \Theta(2\lambda_\phi \mu_\chi^2 - \lambda \mu_\phi^2)$\\
\hline
1 & 1 & 1 & unconstrained & 1 &  unconstrained \\[2mm]
0 & 1 & 1 & 1             & 1 &  unconstrained \\
0 & 1 & 1 & 1             & 0 & 1  \\
\hline
\hline
\end{tabular}
\end{center}
\end{table}

After spontaneous symmetry breaking of $G \to SU(3)_\rc\otimes U(1)_Q$%
\footnote{These are the only gauge symmetries that we could observe in
Nature so far.} we use the following convenient parametrization for the
scalar fields: 
\begin{equation}
\phi=\frac{1}{\sqrt{2}}\,\re^{\ri\bT\cdot\bom{\xi}(x)/v}
\lp(\!\!\begin{array}{c} 0 \\ v+h'(x) \end{array}\!\!\rp)
\quad\mbox{and}\quad
\chi(x) =
\frac{1}{\sqrt{2}}\,\re^{\ri\eta(x)/w}\big(w + s'(x)\big)
\,.
\label{eq:BEHparametrization}
\end{equation}
We can use the gauge invariance of the model to choose the unitary
gauge when
\beq
\phi'(x)=
\frac{1}{\sqrt{2}} \lp(\!\!\begin{array}{c} 0 \\ v+h'(x) \end{array}\!\!\rp)
\quad\mbox{and}\quad
\chi'(x) = \frac{1}{\sqrt{2}}\big(w + s'(x)\big)
\,.
\eeq
With this gauge choice, the scalar kinetic term contains quadratic
terms of the gauge fields from which one can identify mass parameters
of the massive standard model gauge bosons proportional to the
vacuum expectation value $v$ of the BEH field and also that of a
massive vector boson $Z^{'\mu}$ proportional to $w$.

We can diagonalize the mass matrix (quadratic terms) of the two real
scalars ($h'$ and $s'$) by the rotation
\beq
\lp(\!\!\begin{array}{c}
 h \\
 s
\end{array}\!\!\rp) =
\lp(\!\!\begin{array}{cr}
 \cos\tS & -\sin\tS \\
 \sin\tS &  \cos\tS
\end{array}\!\!\rp)
\lp(\!\!\begin{array}{c}
 h' \\
 s'
\end{array}\!\!\rp)
\eeq
where for the scalar mixing angle $\tS \in (-\frac\pi4,\frac\pi4)$ we find
\beq
\sin(2\tS) = - \frac{\lambda v w}
{\sqrt{(\lambda_\phi v^2 - \lambda_\chi w^2)^2 + (\lambda v w)^2}}
\,.
\eeq
The masses of the mass eigenstates $h$ and $s$ are
\beq
M_{h/H} = \lp(\lambda_\phi v^2 + \lambda_\chi w^2
\mp \sqrt{(\lambda_\phi v^2 - \lambda_\chi w^2)^2 + 
(\lambda v w)^2}\rp)^{1/2}
\label{eq:Mhs}
\eeq
where $M_h \leq M_H$ by convention. At this point either $h$ or $H$ can
be the standard model Higgs boson.

As $M_h$ must be positive, the condition
\beq
v^2 w^2 \Big(4 \lambda_\phi \lambda_\chi - \lambda^2\Big) > 0
\label{eq:Mhpositivity}
\eeq
has to be fulfilled. If both VEVs are greater than zero--as needed for
two non-vanishing scalar masses--, then this condition reduces to the
positivity constraint (\ref{eq:positivity}), but with different
meaning. \eqn{eq:positivity} is required to ensure that the potential
be bounded from below if $\lambda<0$, which has to be fulfilled at any
scale. For $\lambda>0$, the potential is bounded from below even
without requiring the constraint (\ref{eq:positivity}).  The inequality
in (\ref{eq:Mhpositivity}) ensures $M_h > 0$, which has to be fulfilled
as long as $v w>0$ independently of the sign of $\lambda$.

The VEV of the BEH field and the mass of the Higgs boson are known
experimentally, $v \simeq 262$\,GeV and $m_H \simeq 131.55$\,GeV
\cite{Buttazzo:2013uya}.  Introducing the abbreviation
$\lSM = \frac12 m_H^2/v^2$, we have $\lSM(\mt) \simeq 0.126$ and we can
distinguish two cases at the weak scale:
(i) $\lambda_\phi(\mt) > \lSM(\mt)$ and
(ii) $\lSM(\mt) > \lambda_\phi(\mt)$.
Then we can relate the new VEV $w$ to the BEH VEV $v$ and the four
couplings $\lSM$, $\lambda_\phi$, $\lambda_\chi$, $\lambda$ using
\eqn{eq:Mhs} as
\beq
w(\mt)^2 (4 (\lambda_\phi(\mt) - \lSM(\mt)) \lambda_\chi(\mt) - \lambda(\mt)^2) =
4 v(\mt)^2 \lSM(\mt) (\lambda_\phi(\mt)-\lSM(\mt))
\,.
\label{eq:w2}
\eeq 
Using \eqn{eq:w2}, it is convenient to consider
$w$ as a dependent parameter and scan the parameter space of the
remaining three quartic couplings as done below. We are not interested
in the case of $\lambda_\phi(\mt)=\lSM(\mt)$ because that prevents
the model from interpreting neutrino masses \cite{Trocsanyi:2018bkm}.

In case (i) when $\lambda_\phi(\mt) > \lSM(\mt)$, then $M_H > m_H$, so only $h$
can be the Higgs particle %($\sin(2\tS)$, $\cos(2\tS) > 0$)
and 
\beq
M_h = m_H
\,,\quad\textrm{while}\quad
M_H = m_H \sqrt{\frac{\lambda_\phi-\lSM}{\lSM}}
\sqrt{\frac{4 \lambda_\phi \lambda_\chi-\lambda^2}
{4 (\lambda_\phi-\lSM) \lambda_\chi -\lambda^2}}
\,.
\label{eq:MhMH1}
\eeq
The positivity of $M_H^2$, in addition to the constraint in
(\ref{eq:Mhpositivity}), also requires that
\beq
4 (\lambda_\phi-\lSM) \lambda_\chi -\lambda^2 > 0
\quad \textrm{or}\quad
\lambda_\phi > \lSM + \frac{\lambda^2}{4\lambda_\chi}
\,.
\label{eq:ci}
\eeq

In case (ii), $m_H^2 > 2 \lambda_\phi v^2 > M_h^2$, so only $H$ can be
the Higgs particle %($\sin(2\tS)$, $\cos(2\tS) < 0$)
and we can express
the masses of the scalars as in \eqn{eq:MhMH1}, with $h$ and $H$
interchanged, or explicitly 
\beq
M_h = m_H \sqrt{\frac{\lSM-\lambda_\phi}{\lSM}}
\sqrt{\frac{4 \lambda_\phi \lambda_\chi-\lambda^2}
{\lambda^2 + 4 (\lSM-\lambda_\phi) \lambda_\chi}}
\quad\textrm{and}\quad
M_H = m_H
\,,
\label{eq:MhMH2}
\eeq
which does not require any further constraint to (\ref{eq:Mhpositivity}).

In principle, it may happen that one of the VEVs vanishes at some
critical scale $t_c$.  In that case, for $t>t_c$ the only scalar
particle is the Higgs boson. Thus, beyond $t_c$ we do not need to
assume the validity of the extra constraints beyond the requirements of
stability and the new scalar sector affects only the RG equations.

\iffalse
We see that the signs in the rotation matrix are tied to the sign of
$\lambda_\phi - \lSM$:
\beq
\sin(2\tS) = \mathrm{sgn}(\lambda_\phi - \lSM) 2\frac{\lambda v w}{M_H^2-M_h^2}
\,,\quad
\cos(2\tS) = \mathrm{sgn}(\lambda_\phi - \lSM) 2\frac{\lambda_\phi v^2-\lambda_\chi w^2}{M_H^2-M_h^2}
\,.
\eeq
\fi

%%%%%%%%%%%%%%%%%%%%%%%%%%%%%%%%%%%%%%%%%%%%%%%%%%%%%%%%%%%%%%%%%%%%%%
%%%%%%%%%%%%%%%%%%%%%%%%%%%%%%%%%%%%%%%%%%%%%%%%%%%%%%%%%%%%%%%%%%%%%%
%%%
%%% Neutrino Yukawa
%%%
%%%%%%%%%%%%%%%%%%%%%%%%%%%%%%%%%%%%%%%%%%%%%%%%%%%%%%%%%%%%%%%%%%%%%%
%%%%%%%%%%%%%%%%%%%%%%%%%%%%%%%%%%%%%%%%%%%%%%%%%%%%%%%%%%%%%%%%%%%%%%

Neutrino oscillation experiments prove that neutrinos have masses,
which in a usual gauge field theoretical description necessitates the
assumption that right handed neutrinos exist.  The existence of the new
scalar allows for gauge invariant Majorana-type Yukawa terms of
dimension four operators for the neutrinos
\beq
\cL^\nu_{\rY} =
- \frac12 \sum_{i,j}
\overline{\nu_{i,\rR}^c}\,(c_\rR)_{ij}\, \nu_{j,\rR}\,\chi
+ {\rm h.c.}
\label{eq:nuYukawa}
\eeq
provided the superscript $c$ denotes the charge conjugate of the field.
%$\nu^c = -\ri \gamma_2 \nu^*$ and the $Z$-charge of the right-handed
%neutrinos and the new scalar satisfy the relation $z_\chi = -2 z_{\nu_\rR}$.
The Yukawa coupling matrix $(c_\rR)_{ij}$
is a real symmetric matrix whose values are not constrained. There are
other gauge invariant Yukawa terms involving the left-handed neutrinos
(see \Ref{Trocsanyi:2018bkm} where all possible terms are taken into account
for neutrino mass generation), but those must contain small Yukawa
couplings, otherwise the left-handed neutrino masses would violate
experimental constraints. In our analysis below we assume that at least
one element of the diagonal matrix $O\,c_\rR\,O^T$, with $O$ being a
suitable orthogonal matrix, can take any value in the range $(0,1)$. We
denote this element by $c_\nu$ below.  

%%%%%%%%%%%%%%%%%%%%%%%%%%%%%%%%%%%%%%%%%%%%%%%%%%%%%%%%%%%%%%%%%%%%%%
%%%%%%%%%%%%%%%%%%%%%%%%%%%%%%%%%%%%%%%%%%%%%%%%%%%%%%%%%%%%%%%%%%%%%%
%%%
%%% Renormalization group equations
%%%
%%%%%%%%%%%%%%%%%%%%%%%%%%%%%%%%%%%%%%%%%%%%%%%%%%%%%%%%%%%%%%%%%%%%%%
%%%%%%%%%%%%%%%%%%%%%%%%%%%%%%%%%%%%%%%%%%%%%%%%%%%%%%%%%%%%%%%%%%%%%%

The values of the couplings at any scale are determined by the RG
equations,
\beq
\label{eq:RGequations}
\frac{\rd g}{\rd t} = a \beta(g)
\eeq
where the factor $a=\ln 10$ ensures, that the RG-time $t=\ln(\mu/\GeV)$
represents the energy scale $\mu=10^t$\,GeV rather than $\mu=e^t$\,GeV
and the variable $g$ is a generic notation for the five gauge couplings,
the four most relevant Yukawa couplings $c_\rt$, $c_\rb$, $c_\tau$ and
$c_\nu$, the two quadratic and three quartic scalar couplings (14
equations in total). In order to solve this coupled system of
differential equations, we need to specify the $\beta$-functions and
the initial conditions for the couplings.

At one-loop in perturbation theory, the $\beta$-function of a
dimensionless coupling $g$ is computed from the formula 
\beq 
\beta_0(g) = M \frac{\partial}{\partial M}
\biggl( -\delta_g + \frac{1}{2} g\sum_i \delta_{Z,i}\biggr)
\label{eq:betadef}
\eeq
where $\delta_g$ is the one-loop counterterm for a given vertex, which
is proportional to $g$, while $\delta_{Z,i}$ are the wave function
renormalization counterterms for all the $i$ legs of the given vertex.
The one-loop $\beta$-functions are scheme independent and so is the
one-loop equation \eqn{eq:betadef} (see for istance Chapter 12.~of
\Ref{Peskin:2007}).  We computed those in perturbation theory at
one-loop order for the complete model of \Ref{Trocsanyi:2018bkm}. For the
sake of completeness, we list those in \app{sec:appendix1}.%
\footnote{It is easy to convince ourselves that the $\beta$-functions
of the scalar sector should not depend on the $Z$-charges. Indeed, our
$\beta$-functions almost coincide with those of \Ref{Basso:2011hn}
written for the $U(1)_{B-L}$ extension, with obvious changes due to
the absence of scalar-vector coupling there.} 
In order to obtain the running of the scalar couplings, we need the
$\beta$-functions of the scalar sector.  According to our assumption on
the smallness of the new gauge couplings, we can set $g'_Z=g'_{ZY}=0$.
We also neglect the Yukawa couplings of all charged leptons as well as
the quarks, except that of the t-quark. With these assumptions the
$\beta$-functions $\beta_0(g) \equiv b_0(g)/(4\pi)^2$ of the gauge
and Yukawa couplings simplify to their forms in the standard model,
while those in the scalar sector become
\beq
\bsp 
%\beta_0(\mu_\phi^2) &= \frac{\mu_\phi^2}{(4\pi)^2}
b_0(\mu_\phi^2) &= \mu_\phi^2
\left(12\lambda_\phi + 2 \frac{\mu_\chi^2}{\mu_\phi^2} \lambda  + 6 c_\rt^2
- \frac{3}{2} g_Y^2  - \frac{9}{2} g_\rL^2 \right)
\,,
\\
%\beta(\mu_\chi^2) &= \frac{\mu_\chi^2}{(4\pi)^2}
b_0(\mu_\chi^2) &=
\mu_\chi^2
\left(8\lambda_\chi + 4 \frac{\mu_\phi^2}{\mu_\chi^2} \lambda + \frac{1}{2} c_\nu^2  \right)
\quad\textrm{with Dirac neutrino}
\,,
\\ &=
\mu_\chi^2
\left(8\lambda_\chi + 4 \frac{\mu_\phi^2}{\mu_\chi^2} \lambda + c_\nu^2  \right)
\quad\textrm{with Majorana neutrino}
\,,
\\
%\beta(\lambda_\phi) &= \frac{1}{(4\pi)^2}\left[
b_0(\lambda_\phi) &=
24\lambda_\phi^2 + \lambda^2  - 6 c_\rt^4 +
\frac{3}{8}g_Y^4 + \frac{9}{8} g_\rL^4 + \frac{3}{4}g_Y^2 g_\rL^2
-\lambda_\phi \bigg(  9 g_\rL^2 + 3 g_Y^2 \bigg) +12 \lambda_\phi  c_\rt^2
%\right]
\,,
\\
%\beta(\lambda_\chi) &= \frac{1}{(4\pi)^2}\left[
b_0(\lambda_\chi) &=
20\lambda_\chi^2 + 2\lambda^2 - \frac{1}{8} c_\nu^4   + \lambda_\chi c_\nu^2
\quad\textrm{with Dirac neutrino}
\,,
\\ &=
20\lambda_\chi^2 + 2\lambda^2 - \frac{1}{2} c_\nu^4   + 2 \lambda_\chi c_\nu^2
\quad\textrm{with Majorana neutrino}
%\right]
\,,
\\
%\beta(\lambda) &= \frac{1}{(4\pi)^2} \left[
b_0(\lambda) &= 
12\lambda \lambda_\phi + 8 \lambda \lambda_\chi + 4 \lambda^2 
+ \lambda \bigg( \frac{1}{4} c_\nu^2 +6 c_\rt^2
- \frac{9}{2} g_\rL^2 - \frac{3}{2} g_Y^2 \bigg)
\quad\textrm{with Dirac neutrino}
\,,
\\ &=
12\lambda \lambda_\phi + 8 \lambda \lambda_\chi + 4 \lambda^2 
+ \lambda \bigg( c_\nu^2 +6 c_\rt^2
- \frac{9}{2} g_\rL^2 - \frac{3}{2} g_Y^2 \bigg)
\quad\textrm{with Majorana neutrino}
%\right]
\,.
\label{eq:betafcns}
\esp
\eeq
We solve this system of simplified equations numerically for both types.
Of course, for $c_\nu=0$ the difference between the equations for Dirac
and Majorana neutrinos disappears. For $c_\nu>0$ the qualitative behaviour
of the running couplings is similar for the two types of neutrinos, but
the larger coefficients in front of $c_\nu$ for the Majorana neutrino
results in a stronger effect of the neutrino Yukawa coupling, and
eventually more constrained parameter space.

We fix the initial conditions for the standard model couplings as done
in the two-loop analysis of \Ref{Buttazzo:2013uya} (using the two-loop
$\overline{\text{MS}}$ scheme).
Specifically, we set
\beq
\bsp
g_Y(\mt) &= \sqrt{\frac{3}{5}}\times 0.4626\,,
\quad
g_\rL(\mt)  =  0.648\,,%0.6477\,,
\quad
g_3(\mt)  = 1.167\,,%1.166\,,
\\
\lSM(\mt) &= 0.126\,,%0.1259\,,
\quad
v(\mt) = 262\,\GeV\,,
\quad
c_\rt(\mt) = 0.937\,.% 0.9379\,.
\esp
\eeq
Chosing some initial values of the quartic couplings $\lambda_\phi(\mt)$,
$\lambda_\chi(\mt)$ and $\lambda(\mt)$, we obtain $\mu_\phi(\mt)$ and
$\mu_\chi(\mt)$ according to \eqn{eq:scalarmasses}, with
$w(\mt)=w\Big(\!\lambda_\phi(\mt),\lambda_\chi(\mt),\lambda(\mt),\lSM(\mt)\!\Big)$
determined from \eqn{eq:w2}.

%%%%%%%%%%%%%%%%%%%%%%%%%%%%%%%%%%%%%%%%%%%%%%%%%%%%%%%%%%%%%%%%%%%%%%
%%%%%%%%%%%%%%%%%%%%%%%%%%%%%%%%%%%%%%%%%%%%%%%%%%%%%%%%%%%%%%%%%%%%%%
%%%
%%% Constraints on the scalar couplings
%%%
%%%%%%%%%%%%%%%%%%%%%%%%%%%%%%%%%%%%%%%%%%%%%%%%%%%%%%%%%%%%%%%%%%%%%%
%%%%%%%%%%%%%%%%%%%%%%%%%%%%%%%%%%%%%%%%%%%%%%%%%%%%%%%%%%%%%%%%%%%%%%

In order to constrain the parameter space of the new couplings, spanned
by $\lambda_\phi$, $\lambda_\chi$, $\lambda$ and $c_\nu$, we require
the validity of the conditions of \tab{tab:scalarcouplings}, i.e.~the
stability of the vacuum up to the Planck scale $\mP$. Such studies
have already been presented for various hidden sector (usually singlet
scalar) extensions of the standard model in
\Refs{Gonderinger:2012rd,Khan:2014kba,Alanne:2014bra,DiChiara:2014wha}.
In addition, we also check the validity of the constraints set
by the positivity requirement on the scalar masses (\eqn{eq:ci} for
case (i) and \eqn{eq:Mhpositivity} for case (ii)), from the initial
conditions up to $\mP$, but as long as $w>0$. A similar analysis was 
presented in \Ref{Duch:2015jta}, but with $Z_2$ symmetry assumed on the
new gauge sector. Our analysis is based on the simplest, but complete
(in the sense of renormalizable quantum field theory) extension of the
standard model gauge group described in \Ref{Trocsanyi:2018bkm}. This
model introduces a new force, mediated by a T vector boson and has the
potential of explaining the confirmed experimental observations that
cannot be interpreted within the standard model.

As seen in \eqn{eq:betafcns}, the $\beta$-functions are independent of
both $\mu_\phi$ and $\mu_\chi$, except of course their own
$\beta$-functions, which decouple from the rest. Thus, in the parameter
scan we focus on the four-dimensional parameter subspace of $c_\nu$,
$\lambda_\phi$, $\lambda_\chi$, $\lambda$ by selecting slices at fixed
values of $c_\nu$. In addition to the stability conditions, we also 
require that the couplings remain in the perturbative region that we
defined by
\beq
\lambda_\phi(t) < 4 \pi\,,\quad
\lambda_\chi(t) < 4 \pi\,,\quad
|\lambda(t)| < 4 \pi\,.
\eeq
We have restricted the region of the new VEV to $w<1$\,TeV because
a large value of $w$ is likely to imply large kinetic mixing between
the two $U(1)$ gauge fields \cite{Trocsanyi:2018bkm}, which is not
supported by experiments (see e.g.~\Ref{Alexander:2016aln}).  This
restriction does not influence the allowed regions for the quartic
couplings significantly.  

\begin{figure}
\begin{center}
\includegraphics[width=0.49\textwidth]{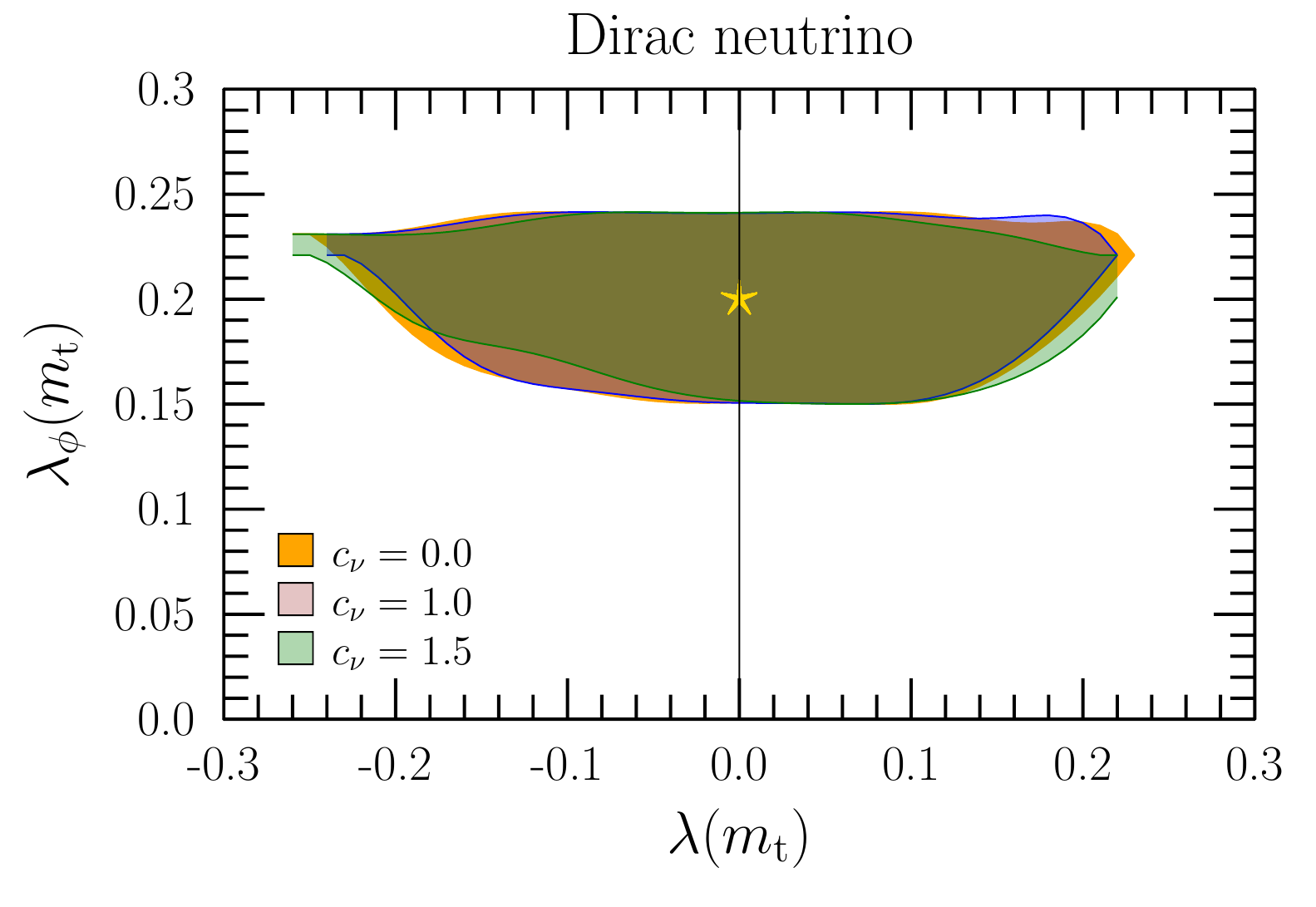}
\includegraphics[width=0.49\textwidth]{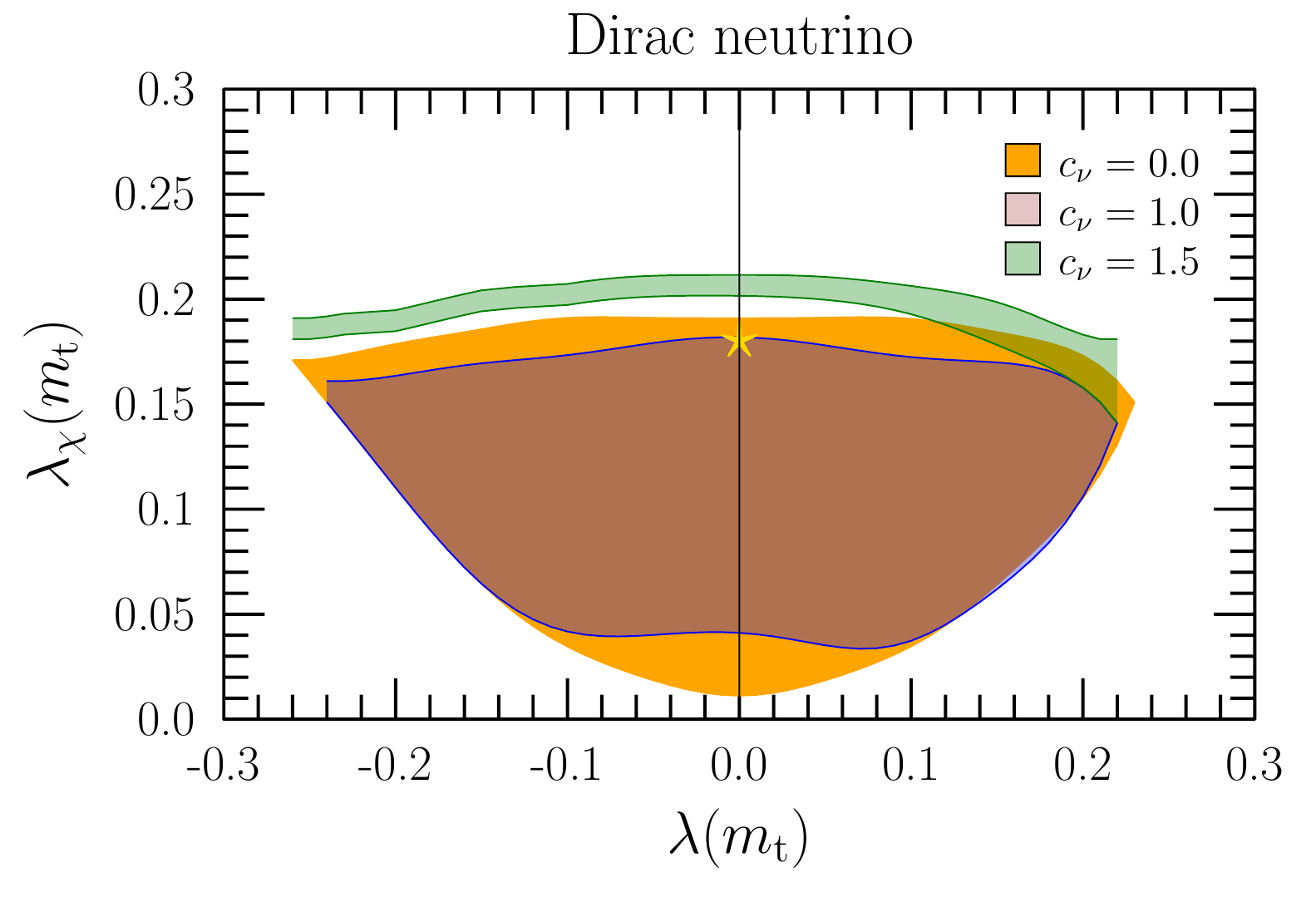}
\end{center}
\caption{\label{fig:1} Accepted initial conditions (as shaded areas)
in the $\lambda_\phi(\mt)-\lambda(\mt)$ plane (left)
and $\lambda_\chi(\mt)-\lambda(\mt)$ plane (right) for the stability
of the vacuum and perturbativity preserved up to the Planck mass
at different values of the Dirac neutrino Yukawa coupling $c_\nu$.
The star marks the point in the parameter space for which the example of
the running couplings up to the Planck scale is presented in \fig{fig:2}} 
\end{figure}

\figs{fig:1}{fig:2} display our results for the allowed regions for
the initial conditions of $\lambda_\phi$, $\lambda_\chi$ and $\lambda$ 
at three selected values of the Dirac neutrino Yukawa
coupling as shaded areas where the stability of the vacuum and the 
constraints set by the positivity requirement on the scalar masses
are respected. In order to ease the interpretation, we show projections
of the allowed region onto two-dimensional subspaces. We also show the 
running couplings up to the Planck scale at a point representing
selected values of the initial conditions at the electroweak scale.
Although the new VEV $w$ is not an independent parameter, we find
interesting to present the projections also in the $w-g$ subspaces
where $g$ denotes one of the quartic couplings. The foremost conclusion
is that the parameter space is not empty, but only for case (i),
i.e.~when $\lambda_\phi(\mt)>\lSM$. Thus the Higgs particle has the
smaller scalar mass always.  In fact, we find that the allowed region
for $\lambda_\phi(\mt)$ is about $[0.151,0.241]$ (starting to decrease
only for $c_\nu(\mt) > 1.5$, while $\min M_H(t) \simeq 144$\,GeV.
Clearly, the precise values may somewhat
change in an analysis at precision of higher loops. Even in the allowed
region for $\lambda_\phi$, the parameter space for the other couplings is
constrained significantly and decreases slowly with increasing Yukawa
coupling of the right handed neutrino up to $c_\nu\simeq 1$. Above
$c_\nu\simeq 1$ the parameter space vanishes swiftly. The maximal
allowed regions for the parameters are presented for the selected
values of $c_\nu$ in \tab{tab:regions}. Thus we find that the stability
of the vacuum requires $c_\nu \lesssim 1.65$ for Dirac neutrinos
($c_\nu\lesssim 1.15$ for Majorana neutrinos).  It is also interesting
to remark that the allowed regions are also very sensitive to the value
of the Yukawa coupling of the t quark. For instance, at $c_\rt(\mt)
\simeq 1.1$ the allowed parameter space vanishes completely.  
\begin{figure}
\begin{center}
\includegraphics[width=0.49\textwidth]{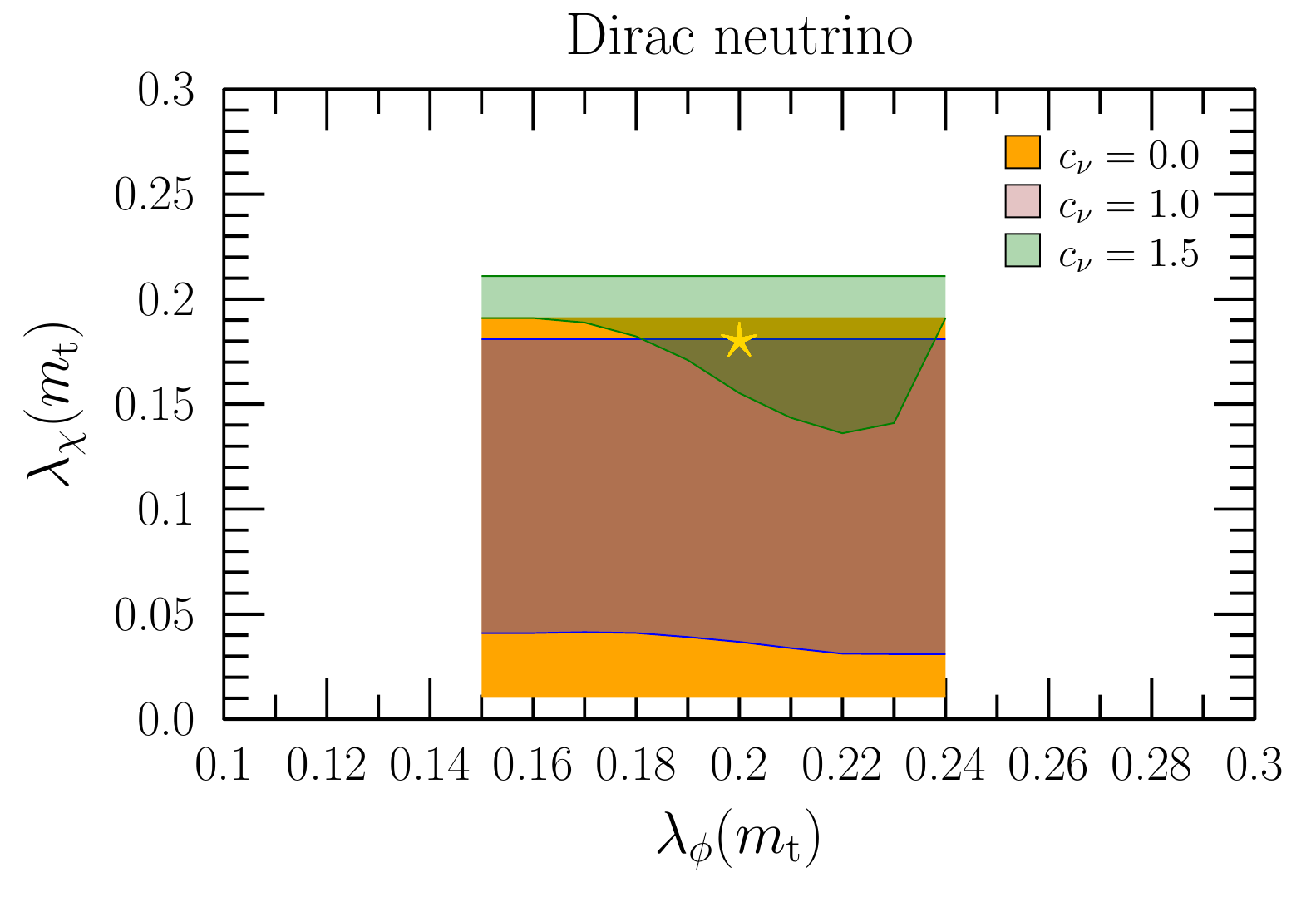}
\includegraphics[width=0.49\textwidth]{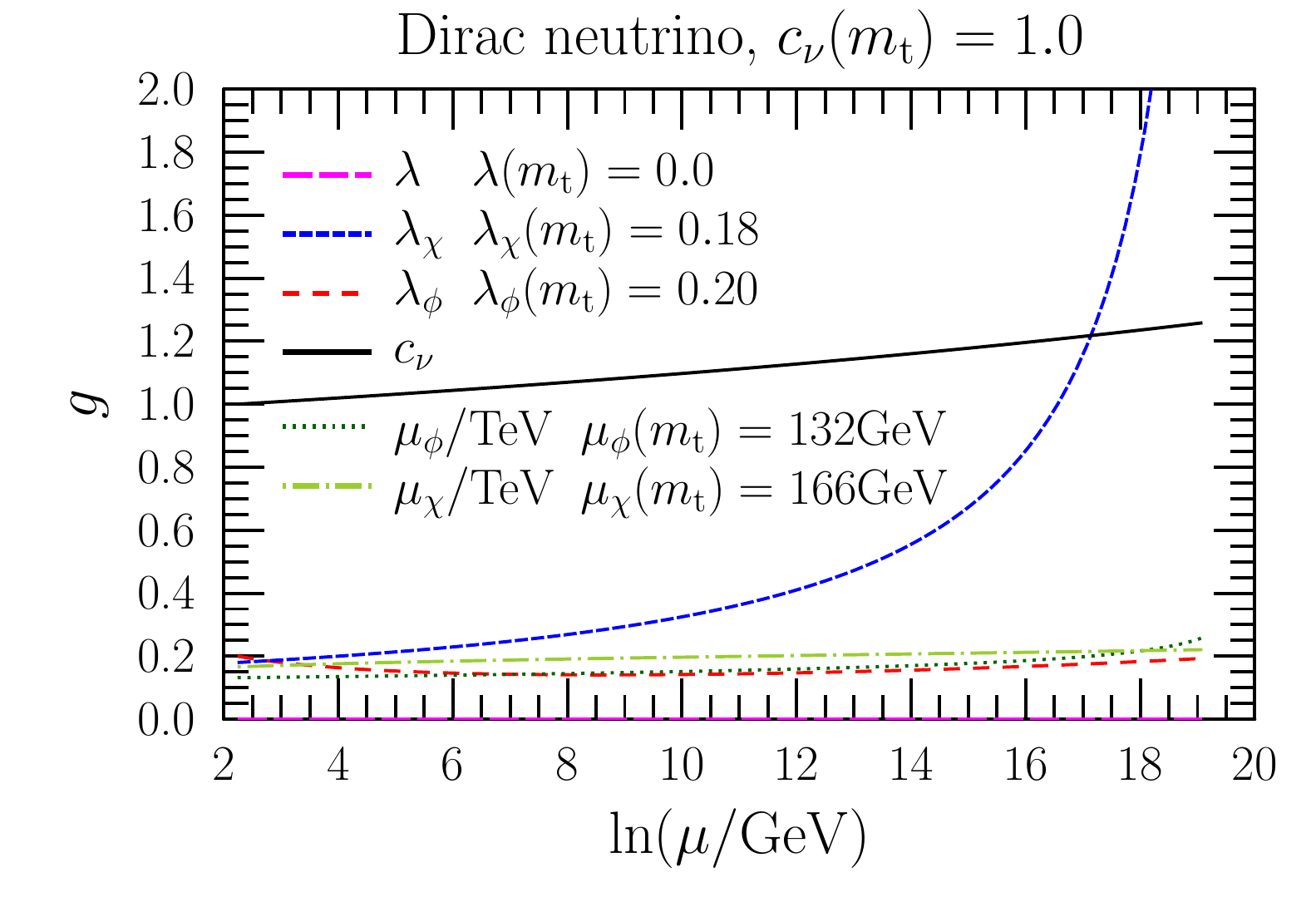}
\end{center}
\caption{\label{fig:2} Left: same as \fig{fig:1} in the
$\lambda_\phi(\mt)-\lambda_\chi(\mt)$ plane.
Right: the running of the couplings up to the Planck scale in a
selected point of the parameter space}
\end{figure}
\begin{figure}
\begin{center}
\includegraphics[width=0.497\textwidth]{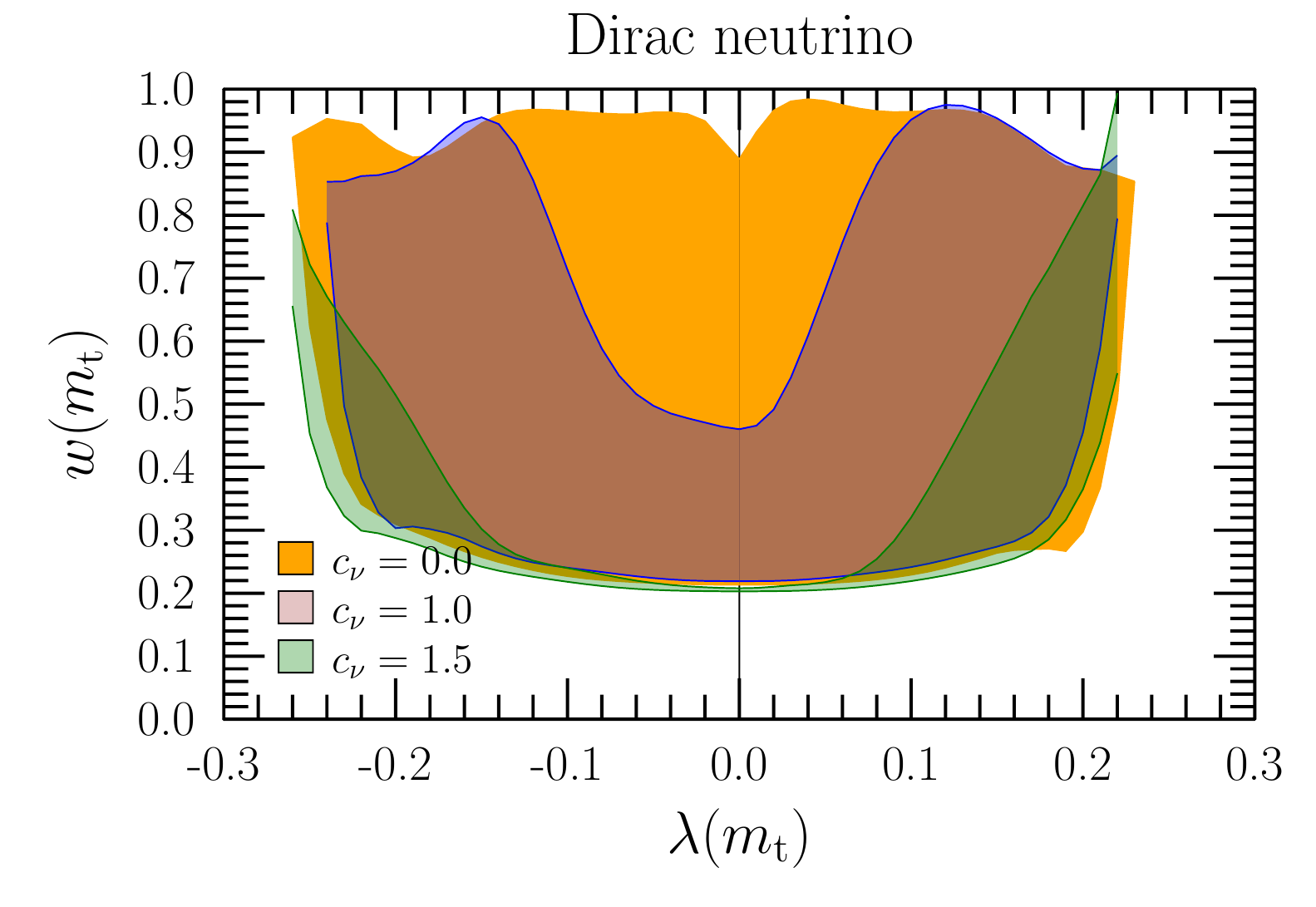}
\includegraphics[width=0.49\textwidth]{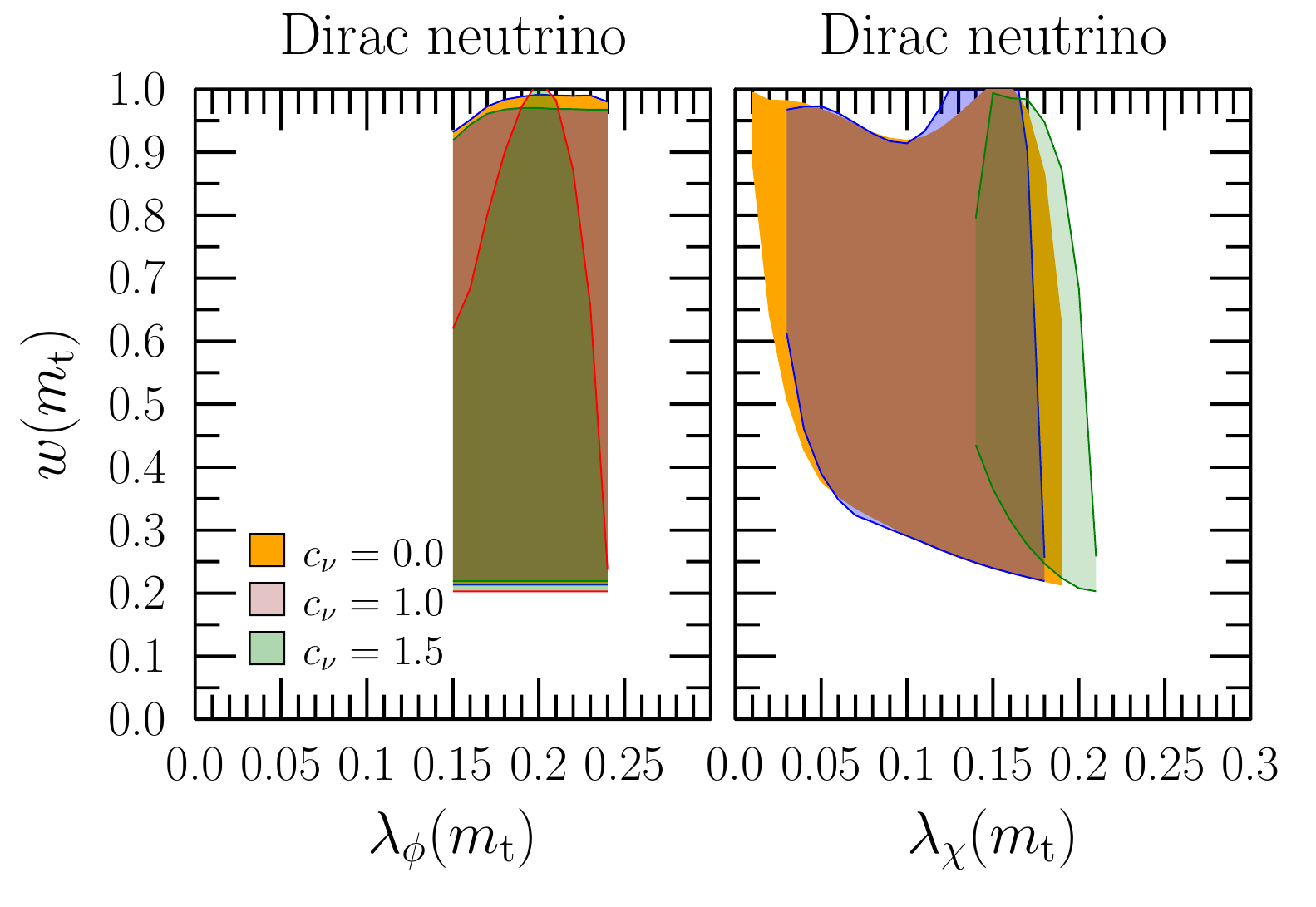}
\end{center}
\caption{\label{fig:3} Same as \fig{fig:1} in the $w(\mt)-g(\mt)$ planes.
Left: $g = \lambda$, right: $g = \lambda_\phi$ and $\lambda_\chi$}
\end{figure}
\begin{table}
\begin{center}
\caption{Maximal allowed regions of the couplings required by stability
of the vacuum and perturbativity of the couplings up to the Planck scale
for selected values of the Yukawa coupling of the right-handed neutrino
$c_\nu(\mt)$ and $w<1$\,TeV set explicitly}
\label{tab:regions}
\begin{tabular}{c|cccccc}
\hline
\hline
$c_\nu$ & $\lambda(t)$ & $\lambda_\chi(t)$ & $\!(\mu_\phi(t)$/GeV$)^2$
& $\!(\mu_\chi^2(t)$/GeV$)^2$ & $\!w(t)$/GeV & $\!M_H(t)$/GeV \\ \hline
0.0 & [-0.26,0.23] & [0.011,0.191] & $[-436^2,496^2]$ & $[112^2,580^2]$ & [213,1000] & [144,558] \\
1.0 & [-0.24,0.22] & [0.031,0.181] & $[-436^2,436^2]$ & $[114^2,548^2]$ & [220,1000] & [144,557] \\
1.5 & [-0.26,0.22] & [0.141,0.211] & $[-374^2,496^2]$ & $[111^2,603^2]$ & [203,994]  & [144,598] \\
%1.6 & [-0.24,0.19] & [0.221,0,231] & & & [205,646] & 436 \\
\hline
\hline
\end{tabular}
\end{center}
\end{table}

%%%%%%%%%%%%%%%%%%%%%%%%%%%%%%%%%%%%%%%%%%%%%%%%%%%%%%%%%%%%%%%%%%%%%%
%%%%%%%%%%%%%%%%%%%%%%%%%%%%%%%%%%%%%%%%%%%%%%%%%%%%%%%%%%%%%%%%%%%%%%
%%%
%%% Conclusions
%%%
%%%%%%%%%%%%%%%%%%%%%%%%%%%%%%%%%%%%%%%%%%%%%%%%%%%%%%%%%%%%%%%%%%%%%%
%%%%%%%%%%%%%%%%%%%%%%%%%%%%%%%%%%%%%%%%%%%%%%%%%%%%%%%%%%%%%%%%%%%%%%

In this letter we studied the ultraviolet behaviour of a simple,
but complete (in the sense of renormalizable quantum field theory)
extension of the standard model gauge group \Ref{Trocsanyi:2018bkm}.
In order to constrain the parameter space of this new model, its
predictions have to be confronted with the large number of established
experimental results in particle physics and cosmology.  We consider
such experimental fact the existence of our Universe, which according
to our assumption, requires the stability of the vacuum up to the
Planck scale. Thus we computed the $\beta$-functions of the model
and studied the dependence of the running couplings of the scalar
sector on the scale. Depending on the initial conditions at low energy
(set at the mass of the t-quark), we find a region in the parameter
space of the new quartic couplings and the largest neutrino Yukawa
coupling where the vacuum remains stable up to the Planck scale.

\subtitle{Acknowledgments}

This work was supported by grant K 125105 of the National Research,
Development and Innovation Fund in Hungary.

%\bibliographystyle{JHEP-2}
%\bibliography{RGanalysis}

\begin{thebibliography}{1}

\bibitem{Weinberg:1967tq}
S.~Weinberg, {\it {A Model of Leptons}},  {\em Phys. Rev. Lett.} {\bf 19}
  (1967) 1264--1266.
%%CITATION = PRLTA,19,1264;%%

\bibitem{ALEPH:2005ab}
{\bf ALEPH, DELPHI, L3, OPAL, SLD, LEP Electroweak Working Group, SLD
  Electroweak Group, SLD Heavy Flavour Group} Collaboration, S.~Schael {\em
  et.~al.}, {\it {Precision electroweak measurements on the $Z$ resonance}},
  {\em Phys. Rept.} {\bf 427} (2006) 257--454
  [\href{http://arXiv.org/abs/hep-ex/0509008}{{\tt hep-ex/0509008}}].
%%CITATION = HEP-EX/0509008;%%

\bibitem{ATLAS}
https://twiki.cern.ch/twiki/bin/view/AtlasPublic/StandardModelPublicResults

\bibitem{CMS}
https://twiki.cern.ch/twiki/bin/view/CMSPublic/PhysicsResultsCombined

\bibitem{Aad:2014aba}
{\bf ATLAS} Collaboration, G.~Aad {\em et.~al.}, {\it {Measurement of the Higgs
  boson mass from the $H\rightarrow \gamma\gamma$ and $H \rightarrow ZZ^{*}
  \rightarrow 4\ell$ channels with the ATLAS detector using 25 fb$^{-1}$ of
  $pp$ collision data}},  {\em Phys. Rev.} {\bf D90} (2014), no.~5 052004
  [\href{http://arXiv.org/abs/1406.3827}{{\tt 1406.3827}}].
%%CITATION = ARXIV:1406.3827;%%

\bibitem{Khachatryan:2014jba}
{\bf CMS} Collaboration, V.~Khachatryan {\em et.~al.}, {\it {Precise
  determination of the mass of the Higgs boson and tests of compatibility of
  its couplings with the standard model predictions using proton collisions at
  7 and 8 $\,\text {TeV}$}},  {\em Eur. Phys. J.} {\bf C75} (2015), no.~5 212
  [\href{http://arXiv.org/abs/1412.8662}{{\tt 1412.8662}}].
%%CITATION = ARXIV:1412.8662;%%

\bibitem{Degrassi:2012ry}
G.~Degrassi, S.~Di~Vita, J.~Elias-Miro, J.~R. Espinosa, G.~F. Giudice,
  G.~Isidori and A.~Strumia, {\it {Higgs mass and vacuum stability in the
  Standard Model at NNLO}},  {\em JHEP} {\bf 08} (2012) 098
  [\href{http://arXiv.org/abs/1205.6497}{{\tt 1205.6497}}].
%%CITATION = ARXIV:1205.6497;%%

\bibitem{Buttazzo:2013uya}
D.~Buttazzo, G.~Degrassi, P.~P. Giardino, G.~F. Giudice, F.~Sala, A.~Salvio and
  A.~Strumia, {\it {Investigating the near-criticality of the Higgs boson}},
  {\em JHEP} {\bf 12} (2013) 089 [\href{http://arXiv.org/abs/1307.3536}{{\tt
  1307.3536}}].
%%CITATION = ARXIV:1307.3536;%%

\bibitem{Trocsanyi:2018bkm}
Z.~Tr\'ocs\'anyi, {\it {Super-weak force and neutrino masses}},
  \href{http://arXiv.org/abs/1812.11189}{{\tt 1812.11189}}.
%%CITATION = ARXIV:1812.11189;%%

\bibitem{Tanabashi:2018oca}
{\bf Particle Data Group} Collaboration, M.~Tanabashi {\em et.~al.}, {\it
  {Review of Particle Physics}},  {\em Phys. Rev.} {\bf D98} (2018), no.~3
  030001.
%%CITATION = PHRVA,D98,030001;%%

\bibitem{delAguila:1995rb}
F.~del Aguila, M.~Masip and M.~Perez-Victoria, {\it {Physical parameters and
  renormalization of U(1)-a x U(1)-b models}},  {\em Nucl. Phys.} {\bf B456}
  (1995) 531--549 [\href{http://arXiv.org/abs/hep-ph/9507455}{{\tt
  hep-ph/9507455}}].
%%CITATION = HEP-PH/9507455;%%

\bibitem{Peskin:2007} M. E. Peskin, D. V. Schroeder, An introduction to quantum
field theory. ABP - The advanced book program. Westview Press, Boulder, Colo.
[u.a.], [reprint] edition, [ca. 2007].
%

\bibitem{Basso:2011hn}
L.~Basso, {\em {Phenomenology of the minimal B-L extension of the Standard
  Model at the LHC}}.
\newblock PhD thesis, Southampton U., 2011.
\newblock \href{http://arXiv.org/abs/1106.4462}{{\tt 1106.4462}}.
%%CITATION = ARXIV:1106.4462;%%

\bibitem{Gonderinger:2012rd}
M.~Gonderinger, H.~Lim and M.~J. Ramsey-Musolf, {\it {Complex Scalar Singlet
  Dark Matter: Vacuum Stability and Phenomenology}},  {\em Phys. Rev.} {\bf
  D86} (2012) 043511 [\href{http://arXiv.org/abs/1202.1316}{{\tt 1202.1316}}].
%%CITATION = ARXIV:1202.1316;%%

\bibitem{Khan:2014kba}
N.~Khan and S.~Rakshit, {\it {Study of electroweak vacuum metastability with a
  singlet scalar dark matter}},  {\em Phys. Rev.} {\bf D90} (2014), no.~11
  113008 [\href{http://arXiv.org/abs/1407.6015}{{\tt 1407.6015}}].
%%CITATION = ARXIV:1407.6015;%%

\bibitem{Alanne:2014bra}
T.~Alanne, K.~Tuominen and V.~Vaskonen, {\it {Strong phase transition, dark
  matter and vacuum stability from simple hidden sectors}},  {\em Nucl. Phys.}
  {\bf B889} (2014) 692--711 [\href{http://arXiv.org/abs/1407.0688}{{\tt
  1407.0688}}].
%%CITATION = ARXIV:1407.0688;%%

\bibitem{DiChiara:2014wha}
S.~Di~Chiara, V.~Keus and O.~Lebedev, {\it {Stabilizing the Higgs potential
  with a Z$'$}},  {\em Phys. Lett.} {\bf B744} (2015) 59--66
  [\href{http://arXiv.org/abs/1412.7036}{{\tt 1412.7036}}].
%%CITATION = ARXIV:1412.7036;%%

\bibitem{Duch:2015jta}
M.~Duch, B.~Grzadkowski and M.~McGarrie, {\it {A stable Higgs portal with
  vector dark matter}},  {\em JHEP} {\bf 09} (2015) 162
  [\href{http://arXiv.org/abs/1506.08805}{{\tt 1506.08805}}].
%%CITATION = ARXIV:1506.08805;%%

\bibitem{Alexander:2016aln} 
J.~Alexander {\it et al.},
{\it {Dark Sectors 2016 Workshop: Community Report}},
[\href{http://arXiv.org/abs/1608.08632}{{\tt 1608.08632}}].
%%CITATION = ARXIV:1608.08632;%%

\end{thebibliography}

\providecommand{\href}[2]{#2}\begingroup\raggedright\endgroup

\appendix
\section{One-loop $\beta$-functions}
\label{sec:appendix1}

We list here the one-loop $\beta$-functions $\beta_0 \equiv
b_0/(4\pi)^2$ of the $U(1)$-extension of \Ref{Trocsanyi:2018bkm} with
the scalar potential in \eqn{eq:V} and for both Dirac 
neutrinos. If the neutrinos are Majorana type, only the coefficients
of the neutrino Yukawa coupling change, which can be found explicitly
in \eqn{eq:betafcns}. For the $U(1)$-gauge couplings we have
\beq 
\bsp
b_0(g_Y) &= \frac{41}{6}g_Y^3\,,
\\
b_0(g'_Z) &=
g'_Z\biggl(\frac{41}{6}g'_{ZY}{}^2
+18g'_Z{}^2+\frac{50}{3}g'_Z g'_{ZY} \biggr)\,,
\\
b_0(g'_{ZY}) &=
g'_{ZY}\biggl(\frac{41}{6}g'_{ZY}{}^2
+18g'_Z{}^2+\frac{50}{3}g'_Z g'_{ZY}\biggr)\,,
\esp
\eeq
without using the GUT normalization.
The $\beta$-functions of the weak and strong couplings at
one-loop level remain the same as in the standard model:
\beq
b_0(g_\rL) =  -\frac{19}{6}g_\rL^3\,,
\qquad
b_0(g_3) =  -7g_3^3\,.
\eeq
The $\beta$-functions for the Yukawa-couplings are:
\beq
\bsp
b_0(c_\nu)
&=
c_\nu\biggl(\frac34 c_\nu^2 - 6 g'_Z{}^2 \biggr)
\\
b_0(c_\tau) 
&= 
c_\tau\biggl( \frac{5}{2}c_\tau^2 + 3 c_\rt^2 + 3 c_\rb^2 - \frac{9}{4}g_\rL^2
- \frac{15}{4}g_Y^2 - \frac{15}{4} g'_{ZY}{}^2 - 3 g'_Z{}^2 - 7 g'_Z g'_{ZY} \biggr)
\,,\\
b_0(c_\rt) 
&= 
c_\rt\biggl( \frac{9}{2}c_\rt^2 + \frac{3}{2} c_\rb^2 +  c_\tau^2 - 8 g_3^2 -
\frac{9}{4}g_\rL^2 - \frac{17}{12}g_Y^2 - \frac{17}{12} g'_{ZY}{}^2
- 3 g'_Z{}^2 - 3 g'_Z g'_{ZY} \biggr)
\,,\\
b_0(c_\rb) 
&= 
c_\rb\biggl( \frac{9}{2}c_\rb^2 + \frac{3}{2} c_\rt^2 +  c_\tau^2 - 8 g_3^2 -
\frac{9}{4}g_\rL^2 - \frac{5}{12}g_Y^2 - \frac{5}{12} g'_{ZY}{}^2 - 3 g'_Z{}^2
- 3 g'_Z g'_{ZY} \biggr)\,.
\esp
\eeq
The scalar mass terms exhibit RG-evolution according to:
\beq 
b_0(\mu_\phi^2) 
=
\mu_\phi^2\biggl(12\lambda_\phi + 2\frac{\mu_\chi^2}{\mu_\phi^2} \lambda
+ 2 c_\tau^2 + 6 c_\rb^2 + 6 c_\rt^2 - \frac{3}{2} g_Y^2  - \frac{9}{2} g_\rL^2 
- 6 g'_Z{}^2 - \frac{3}{2}g_{ZY}^2 -6 g'_Z g'_{ZY} \biggr)
\eeq
and
\beq
b_0(\mu_\chi^2) =
\mu_\chi^2\biggl(8\lambda_\chi + 4 \frac{\mu_\phi^2}{\mu_\chi^2} \lambda
+ \frac12 c_\nu^2  - 24 g'_Z{}^2 \biggr)
\eeq
Finally, the $\beta$-functions for the scalar quartic couplings are
\beq 
\bsp
b_0(\lambda_\phi) &= 
24\lambda_\phi^2 + \lambda^2 - 2 c_\tau^4 - 6 c_\rt^4 - 6 c_\rb^4
+ \frac{3}{8}g_Y^4 + \frac{9}{8} g_\rL^4 + \frac{3}{4}g_Y^2 g_\rL^2
\\
& + 6\lp(g'_Z + \frac{g'_{ZY}}{2}\rp)^4  + (g_Y^2 + g_\rL^2)\lp(g'_Z + \frac{g'_{ZY}}{2}\rp)^2 
\\
&-\lambda_\phi \biggl[9 g_\rL^2 + 3 g_Y^2
+ 12 \lp(g'_Z + \frac{g'_{ZY}}{2}\rp)^2\biggr]
+ 4 \lambda_\phi \biggl(c_\tau^2 +3 c_\rt^2 + 3 c_\rb^2 \biggr)
\,,
\esp
\eeq
\beq 
b_0(\lambda_\chi) = 
20\lambda_\chi^2 + 2\lambda^2 - \frac18 c_\nu^4  + 96 g'_Z{}^2
+ \lambda_\chi c_\nu^2 - 24 \lambda_\chi g'_Z{}^2 
\eeq
and
\beq 
\bsp
b_0(\lambda) &=
12\lambda \lambda_\phi + 8 \lambda \lambda_\chi + 4 \lambda^2 
+ 48 g'_Z{}^2\lp(g'_Z + \frac{g'_{ZY}}{2}\rp)^2 + 24 g_Y^2 g'_Z{}^2 
\\
&
- \lambda
\biggl[8 g'_Z{}^2 + \frac{9}{2} g_\rL^2 + \frac{3}{2} g_Y^2
+ 4\lp(g'_Z  + \frac{g'_{ZY}}{2}\rp)^2\biggr]
\\
&+2\lambda \biggl(\frac14 c_\nu^2 + c_\tau^2 + 3 c_\rt^2 + 3 c_\rb^2 \biggr)
\esp
\eeq

\end{document}